# Quantum Confined Tomonaga-Luttinger Liquid in MoSe$_2$ Twin Domain Boundaries

Yipu Xia,[1] Junqiu Zhang,[1] Yuanjun Jin,[1,2] Wingkin Ho,[1] Hu Xu,[2] and Maohai Xie[1*]

[1]*Physics Department, the University of Hong Kong, Pokfulam Road, Hong Kong*
[2] *Department of Physics, Southern University of Science and Technology, Shenzhen, Guangdong 518055, China*

Abstract:

There have been conflicting reports on the electronic properties of twin domain boundaries (DBs) in MoSe$_2$ monolayer, including the quantum well states, charge density wave, and Tomonaga-Luttinger liquid (TLL). Here we employ low-temperature scanning tunneling spectroscopy to reveal both the quantum confinement effect and signatures of TLL in the one-dimensional DBs. The data do not support the CDW at temperatures down to ~5 K.

Email: mhxie@hku.hk

Recently in the study of monolayer (ML) transition-metal dichalcogenides (TMDs), twin domain boundaries (DBs), especially the so-called 4|4P type commonly seen in epitaxial MoSe$_2$, have drawn much attention [1-4]. These defects behave as metals, being sandwiched between pristine TMD domains and supported by the van der Waals (vdW) substrates, they represent an ideal one-dimensional (1D) system for studying the properties and physics that are pertinent to 1D metals. Over the past decades, a number of 1D systems have been intensively studied, including carbon nanotubes [5-8], semiconductor nanowires [9-11], metal chain adsorbed on semiconductor [12], nanowire bundles [13] and quasi-1D organic conductor [14]. Some interesting properties are expected. For example, they are prone to Peierls instability and may exhibit metal-insulator transition at low-temperature [15-17]. Electron-electron interaction in 1D system invalidates the Landau description of Fermi-liquid and is instead described by the Tomonaga-Luttinger liquid (TLL) theory of collective excitation [18-20]. The latter is characterized by novel properties like spin-charge separation and power-law suppression of electronic density-of-states (DOS) near the Fermi level ($E_F$) [21,22], which have been reportedly observed in various (quasi-)1D systems by transport [9,10,23-25], scanning tunneling spectroscopy [4,12,26,27], and photoemission spectroscopy [3,5,28,29] studies. For the twin DBs in ML MoSe$_2$, TLL was also suggested by a photoemission spectroscopy experiment [3], whereas a low-temperature scanning tunneling microscopy and spectroscopy (LT-STM/S) study pointed to the Peierls type CDW [2]. On the other hand, Liu *et al.* reported observation of quantum well states in finite length DBs as well as an effect of the Moiré potential in system [1]. In a recent paper, Jolie *et al.* provided direct evidence of spin-charge separation, signifying the TLL, but in a different type (i.e., the so called 4|4E type) 1D domain boundary in MoS$_2$ [4].

Stimulated by the above works, we have carried out a systematic LT-STM/S study of the common 4|4P-type DB defects in MoSe$_2$ ML grown by molecular-beam epitaxy (MBE). Dense networks of the 4|4P DBs form in epitaxial MoSe$_2$ and their density is tunable by changing the MBE conditions. As a result, DBs of varying lengths are achieved. By examining the electronic properties of such defects of

various lengths at low temperature (~5 K and 77 K), we show evidence of both the QWS and TLL. The data do not support the CDW in the system.

Fig. 1(a) shows an STM image of an epitaxial MoSe$_2$ ML grown on graphene, revealing a dense DB network with each DB segment 5~10nm long. Surrounding each DB defect are two semiconducting MoSe$_2$ domains having a bandgap of ~2.1 eV as measured by STS [1,30]. The atomic structure of the defect has been well established (see Fig. 1b) and the corresponding electronic structure as calculated from the density functional theory (DFT) is shown in Fig. 1(c) [31-34]. For the latter, we have upshifted the Fermi level to a value that matches with the experiment, i.e., at $k \approx \frac{1}{3}\left(\frac{\pi}{a}\right)$, around one-third of the Brillouin zone (BZ) edge. This procedure had also been adopted in a previous report to explain the CDW in the DBs [2].

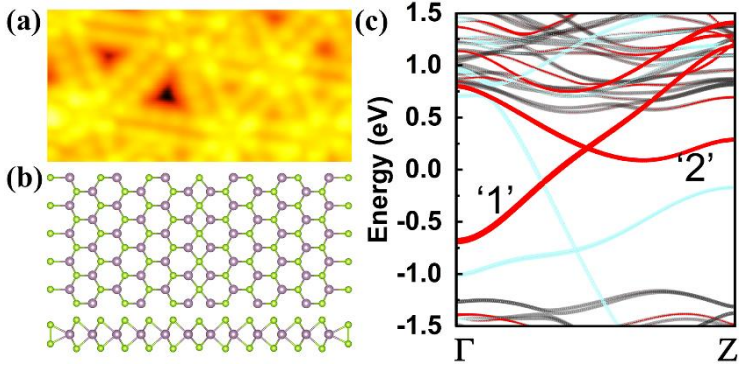

FIG. 1. Atomic and electronic structures of DBs in MoSe$_2$. (a) STM image (15 × 7.5 nm$^2$, sample bias: 0.4V) showing the DBs in MBE-grown MoSe$_2$ ML. (b) Stick-and-ball model of a DB in MoSe$_2$, where purple and green balls represent Mo and Se atoms respectively. (c) DFT calculated band structure, where states of the DB are highlighted in red. Black and light-blue lines are states from the bulk and edge of MoSe$_2$ ML domains, respectively.

In Fig. 2(a), we show a set of STS spectra taken on the DBs of different lengths, revealing indeed narrow gaps at $E_F$. However, one notes that the size of the energy gap varies with the DB length $L$, and Fig. 2(b) summarizes the data that we have collected from a number of defects with lengths ranging from 3.1 nm to 40.7 nm as measured by STS at ~5 K. As seen, the data can be reasonably fit by the inverse function $E_g \sim 1/L$ (solid line), a result that may not be attributed to CDW where a constant, length-independent, gap would be otherwise expected [4]. To further examine the likelihood of CDW formation in the 4|4P DBs in MoSe$_2$, we have performed DFT calculations testing the stability of artificially distorted structures with periodicity 3$a$, where every three unit cells is compressed by 5 pm, 10 pm or 15 pm. It is found that after relaxation, the undistorted structure is returned, suggesting that such distorted lattices are not stable. Therefore, the Peierls distortion of the 4|4P DBs appears less favorable from the energy point of view, thus it may not be the cause of the energy gap opening at $E_F$ as seen in STS. On the other hand, such energy gaps may well be explained by a quantum confinement effect of the TLL, where the zero-mode gap coming mostly from the Coulomb-blockade is inversely proportional to $L$ [4,20].

The QWS in finite length DBs can be better visualized by the color plot of differential conductance $\frac{dI}{dV}$ measured by STS as functions of energy and position along the DB. An example is shown in Fig. 2(c), in which spatial undulation of the charge density as well as energy quantization are clearly discerned. The number of undulation period increases continuously with energy, from 4 (around -0.26eV) to 7 (around +0.18eV), which appears consistent with the calculated defect band '1' as marked in Fig. 1(c). At energy ~0.2 eV, a much-enhanced intensity is noted, which may correspond to the emergence of band '2'. Above 0.2 eV, there are degenerate states due to overlapping bands '1' and '2', complicating the data. In Fig. 2(c), another acute feature is the wide energy gap between states adjacent to $E_F$, much wider than the QWS gaps away from $E_F$. This is precisely the charge-gap of the TLL as described above.

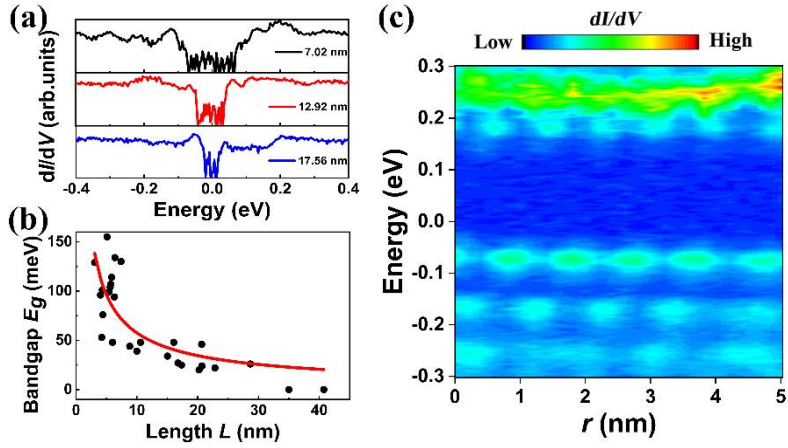

FIG. 2. Quantum well states in finite length DBs. (a) STS spectra ($\frac{dI}{dV}$ on a logarithmic scale) of DBs with three different lengths as indicated. (b) The DOS gap sizes measured by STS of varying DB lengths. The red line represents a lest square fitting by $E_g \sim 1/L$. (c) Spatially-resolved STS spectra taken in the middle region of a DB of total length ~7 nm.

The TLL state in the DBs is further evidenced by the observation of power-law DOS suppression at $E_F$ for defects longer than 30 nm, where quantum effect becomes less significant. Indeed, for such long DBs, we could hardly observe the charge-gap $E_g$. Instead, the measured spectra exhibit power-law suppression of the DOS close to Fermi level, characteristic of the TLL [19,35-38]. As an example, Fig. 3(a) presents a spectrum taken over a DB segment of over 30 nm long (inset), and Fig. 3(b) presents a close-up spectrum near the Fermi energy. As seen, instead of a gap, a power-law suppression of the DOS, $N(E)$, may be observed, and a fit by $N(E) \sim |E|^\alpha$ results in an exponent $\alpha = 0.47 \pm 0.05$. The latter relates to the Luttinger parameter $K_c$ according to

$\alpha = \left(K_c + K_c^{-1} - 2\right)/4$ [20,37,39], which characterizes the strength of electron-electron interaction in the system. From the above, we may derive a value of $K_c \approx 0.28$, which is not too far from that reported in Ref. [4] for the same defect. Fig. 3(c) shows the spatially-resolved STS spectra taken from the middle part of the long DB, visualizing the effective DOS suppression near the Fermi level but reminiscent of the quantum confinement effect. The latter would make the estimate of the parameter $K_c$ shown above less reliable.

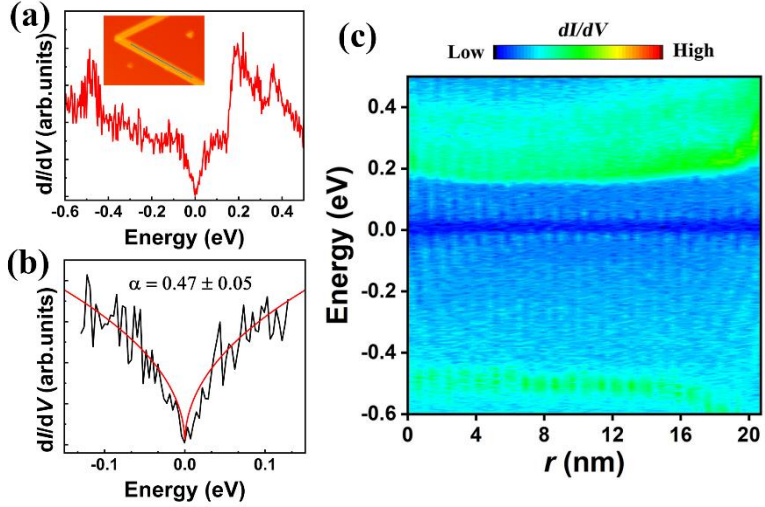

FIG. 3 DOS power-law suppression in long DBs. (a) STS spectra obtained from a DB of total length ~30 nm long shown in the inset (image size: 30 × 21 nm², sample bias: -1.0V). (b) A close-up view of (a) over a narrow energy range. The red line represents a fit of the data by the power-law with an exponent $\alpha = 0.47 \pm 0.05$. (c) Spatially-resolved STS along part of the long DB.

A more direct evidence of TLL may come from spin-charge separation, which could be observable in the STS experiments as demonstrated by Jolie *et al.* [4] Unfortunately our data do not unambiguously show such a feature. On the other hand, we observe another feature that may lend support to the TLL, i.e., increasing suppression of the DOS as one moves closer to the DB edge. According to a theoretical study, the energy at which the DOS peaks shifts to higher values with decreasing *r*, the distance from the 1D wire edge, such that $r \times E$ = constant [37,38]. Fig. 4(a) presents a spatially-resolved STS spectra taken near one end of a DB, on which a curve of $r \times E$ = 0.49 nm·eV is overlaid. Fig. 4(b) plots the same data in a different format, where each line represents a spectrum taken at a location a distance from one of the DB end as indicated. As is clear, the closer it is to the end of the defect, the higher the energy at which the DOS is the highest.

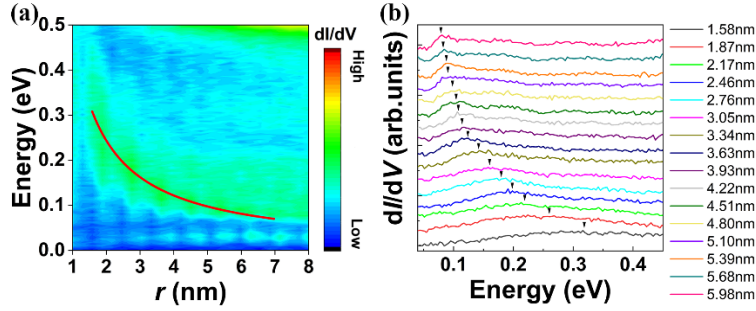

FIG. 4. DOS peak energy as a function of tip position of measurement. (a) Spatially-resolved STS taken near one end (defined as the origin) of a DB, and the red line represents a fit by $E\sim\mathrm{constant}/r$. (b) The $\mathrm{d}I/\mathrm{d}V$ spectra taken at different locations on the DB as indicated, showing the peak energy shift with STM tip position.

In summary, using LT-STM/S, we reveal both the quantum well states and signatures of Tomonago-Luttinger liquid in the 4|4P-type twin domain boundary defect in $MoSe_2$ ML. In particular, we observe the length dependent energy gaps for short DBs but power-law DOS suppression at the Fermi level. In addition, enhancement of DOS suppression as one moves closer to the defect end is observed, and a strong electron interaction in the system is hinted. It shows that the DB defect in $MoSe_2$ ML could be an ideal isolated 1D system for studying low-dimensional and correlated physics.

Acknowledgement:

We are benefitted from discussions with C.J. Wang. The work is supported by a Collaborative Research Grant (No. C7036-17W) from the Research Grant Council of Hong Kong Special Administrative Region, China.

Y. X. and J. Z. contributed equally to this work.

Reference:
[1] H. Liu *et al.*, Physical Review Letters **113**, 066105 (2014).
[2] S. Barja *et al.*, Nature Physics **12**, 751 (2016).
[3] Y. Ma *et al.*, Nature communications **8**, 14231 (2017).
[4] W. Jolie *et al.*, Physical Review X **9**, 011055 (2019).
[5] H. Ishii *et al.*, Nature **426**, 540 (2003).
[6] R. Egger and A. O. Gogolin, Physical review letters **79**, 5082 (1997).
[7] C. Kane, L. Balents, and M. P. Fisher, Physical review letters **79**, 5086 (1997).
[8] Z. Yao, H. W. C. Postma, L. Balents, and C. Dekker, Nature **402**, 273 (1999).
[9] O. Auslaender, H. Steinberg, A. Yacoby, Y. Tserkovnyak, B. Halperin, K. Baldwin, L. Pfeiffer, and K. West, Science **308**, 88 (2005).
[10] Y. Jompol, C. Ford, J. Griffiths, I. Farrer, G. Jones, D. Anderson, D. Ritchie, T. Silk, and A. Schofield, Science **325**, 597 (2009).
[11] A. Yacoby, H. Stormer, N. S. Wingreen, L. Pfeiffer, K. Baldwin, and K. West, Physical review letters **77**, 4612 (1996).
[12] C. Blumenstein *et al.*, Nature Physics **7**, 776 (2011).
[13] L. Venkataraman, Y. S. Hong, and P. Kim, Phys Rev Lett **96**, 076601 (2006).
[14] R. Claessen, M. Sing, U. Schwingenschlögl, P. Blaha, M. Dressel, and C. S. Jacobsen, Physical

review letters **88**, 096402 (2002).

[15] H. W. Yeom *et al.*, Physical review letters **82**, 4898 (1999).

[16] J. Ahn, J. Byun, H. Koh, E. Rotenberg, S. Kevan, and H. Yeom, Physical review letters **93**, 106401 (2004).

[17] J. R. Ahn, H. W. Yeom, H. S. Yoon, and I. W. Lyo, Physical Review Letters **91**, 196403 (2003).

[18] S.-i. Tomonaga, in *Bosonization* (World Scientific, 1994), pp. 63.

[19] J. Luttinger, Physical Review **119**, 1153 (1960).

[20] J. Voit, Reports on Progress in Physics **58**, 977 (1995).

[21] A. E. Mattsson, S. Eggert, and H. Johannesson, Phys Rev B **56**, 15615 (1997).

[22] C. L. Kane and M. P. Fisher, Phys Rev Lett **68**, 1220 (1992).

[23] M. Bockrath, D. H. Cobden, J. Lu, A. G. Rinzler, R. E. Smalley, L. Balents, and P. L. McEuen, Nature **397**, 598 (1999).

[24] B. Gao, A. Komnik, R. Egger, D. C. Glattli, and A. Bachtold, Phys Rev Lett **92**, 216804 (2004).

[25] M. Hashisaka, N. Hiyama, T. Akiho, K. Muraki, and T. Fujisawa, Nature Physics **13**, 559 (2017).

[26] J. Hager, R. Matzdorf, J. He, R. Jin, D. Mandrus, M. Cazalilla, and E. W. Plummer, Physical review letters **95**, 186402 (2005).

[27] F. Wang, J. Alvarez, S.-K. Mo, J. Allen, G.-H. Gweon, J. He, R. Jin, D. Mandrus, and H. Höchst, Physical review letters **96**, 196403 (2006).

[28] P. Segovia, D. Purdie, M. Hengsberger, and Y. Baer, Nature **402**, 504 (1999).

[29] Y. Ohtsubo *et al.*, Physical review letters **115**, 256404 (2015).

[30] Y. Xia *et al.*, 2D Materials **5**, 041005 (2018).

[31] Y. Zhang *et al.*, Nature nanotechnology **9**, 111 (2014).

[32] T. Böker, R. Severin, A. Müller, C. Janowitz, R. Manzke, D. Voß, P. Krüger, A. Mazur, and J. Pollmann, Phys Rev B **64**, 235305 (2001).

[33] J. Hong *et al.*, Nano Letters **17**, 6653 (2017).

[34] J. Lin, S. T. Pantelides, and W. Zhou, Acs Nano **9**, 5189 (2015).

[35] V. Meden, W. Metzner, U. Schollwöck, O. Schneider, T. Stauber, and K. Schönhammer, The European Physical Journal B-Condensed Matter and Complex Systems **16**, 631 (2000).

[36] J. Voit, Journal of Physics: Condensed Matter **5**, 8305 (1993).

[37] S. Eggert, Phys Rev Lett **84**, 4413 (2000).

[38] J. Lee, S. Eggert, H. Kim, S.-J. Kahng, H. Shinohara, and Y. Kuk, Physical review letters **93**, 166403 (2004).

[39] H. Schulz, Physical review letters **64**, 2831 (1990).